\documentclass[
  journal=chr,
  manuscript=article-type,
  year=2020,
  volume=37,
]{cup-journal}

\usepackage{xspace}
\usepackage{amsmath}
\usepackage[nopatch]{microtype}
\usepackage{booktabs}
\usepackage{multirow}
\usepackage{listings}
\usepackage{tcolorbox}

\definecolor{mylightblue}{HTML}{e7f1f8}
\definecolor{mydarkblue}{HTML}{a4ccec}
\tcbuselibrary{listingsutf8} % For Unicode characters, if needed
\tcbset{
    promptstyle/.style={
        colframe=mydarkblue,    % Frame color
        colback=mylightblue,            % Background color
        coltitle=black,            % Title text color
        fonttitle=\small\bfseries,       % Title font
        fontupper=\scriptsize\ttfamily,       % Content font
        rounded corners,           % Rounded corners for the box
        boxrule=0.4mm,             % Border thickness
        toptitle=2mm,              % Space above the title
        bottomtitle=2mm,           % Space below the title
        width=\textwidth,          % Full text width
        enlarge left by=0mm,       % Box padding left
        enlarge right by=0mm,      % Box padding right
    }
}

\newtcolorbox{promptbox}[1][]{
    promptstyle,
    title=Prompt Example, % Default title
    #1                     % Allow for custom title input
}

\newcommand{\llm}{LLM\xspace}
\newcommand{\llms}{LLMs\xspace}

\title{LLM Agents for Interactive Exploration of Historical Cadastre Data: Framework and Application to Venice}

\author{T. Karch}
\affiliation{DH-Lab, EPFL, Lausanne, Switzerland}
\alsoaffiliation{Joint first authors}
\email[F. Author]{tristan.karch@gmail.com}

\author{J. Saydaliev}
\affiliation{DH-Lab, EPFL, Lausanne, Switzerland}
\alsoaffiliation{Joint first authors}

\author{I. Di Lenardo}
\affiliation{DH-Lab, EPFL, Lausanne, Switzerland}

\author{F. Kaplan}
\affiliation{DH-Lab, EPFL, Lausanne, Switzerland}


\begin{document}

\begin{abstract}
Cadastral data reveal key information about the historical organization of cities but are often non-standardized due to diverse formats and human annotations, complicating large-scale analysis. We explore as a case study Venice's urban history during the critical period from 1740 to 1808, capturing the transition following the fall of the ancient Republic and the Ancien Régime. This era's complex cadastral data, marked by its volume and lack of uniform structure, presents unique challenges that our approach adeptly navigates, enabling us to generate spatial queries that bridge past and present urban landscapes. We present a text-to-programs framework that leverages Large Language Models (\llms) to process natural language queries as executable code for analyzing historical cadastral records. Our methodology implements two complementary techniques: a SQL agent for handling structured queries about specific cadastral information, and a coding agent for complex analytical operations requiring custom data manipulation. We propose a taxonomy that classifies historical research questions based on their complexity and analytical requirements, mapping them to the most appropriate technical approach. This framework is supported by an investigation into the execution consistency of the system, alongside a qualitative analysis of the answers it produces. By ensuring interpretability and minimizing hallucination through verifiable program outputs, we demonstrate the system's effectiveness in reconstructing past population information, property features, and spatiotemporal comparisons in Venice.
\end{abstract}

% \section*{Impact Statement}

% Some Cambridge Journals (such as our Prisms journals) require an Impact Statement section. Comment out this section if it is not required.

\section*{Plain Language Summary}

This study applies Large Language Models (LLMs) to historical cadastral data, focusing on Venice from 1740 to 1808, a period of major political and social change. Cadastral records, while valuable for urban history, are often unstandardized and challenging to analyze at scale. To address this, a framework is introduced that converts natural language queries into executable code, enabling structured database searches (SQL) and complex analytical operations (Python). By categorizing historical research questions based on complexity, the methodology ensures accuracy and minimizes errors. This approach reconstructs past urban landscapes, uncovering insights into property ownership, land use, and socio-economic patterns. The findings highlight LLMs’ potential for historical urban studies, offering a scalable and reliable method applicable across disciplines.

\section{Introduction}

\subsection{Context}

Historical cadastral records, widely distributed throughout Europe, serve as invaluable documents to reconstruct past urban and territorial information \citep{kain_cadastral_1992}. These records document property ownership, usage functions, and other essential elements for taxation, offering high confidence in their reliability due to their administrative purpose \citep{bloch_plans_1929}. Often paired with cartographic mappings, these dual systems combine textual descriptions with geographic representations following standardized visual and ontological codes to minimize subjective interpretation and enhance utility for taxation \citep{bourguet_dechiffrer_1988}. The evolution of cadastral sources has been extensively studied, with analyses spanning specific case studies to comparative frameworks. Associated cartography, particularly before and after the Napoleonic introduction of geometric-parcel cadastres, reflects a shift towards standardized cartographic practices \citep{clergeot_cent_2007}. These sources are critical for reconstructing historical population data, property functions, and spatial dynamics, and offer rich opportunities for analyzing urban development and socio-economic structures.\\

\noindent While digitization has increased access to cadastral records, efficiently querying and analyzing them remains a significant challenge. Traditional methods—based on close reading and manual data extraction—are labor-intensive and difficult to scale across time and space. As researchers seek to engage with these sources at scale, there is a growing need for computational approaches that preserve historical nuance while enabling forms of distant reading. This also aligns with recent calls for more design-oriented methodologies in cadastral research, which emphasize the development of flexible tools for information processing~\citep{CAGDAS201177}.

\subsection{Objectives and Contributions}

\begin{figure}[!h]
    \centering
    \includegraphics[width=.85\linewidth]{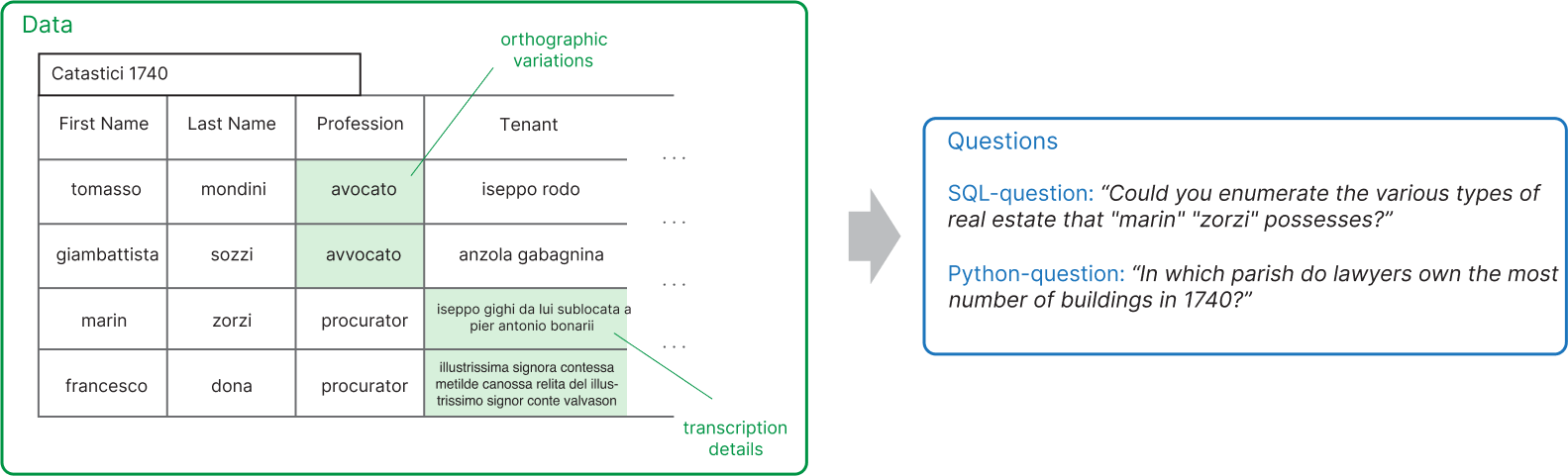}
    \caption{\textbf{The role of \llms in processing historical Cadastres.} Processing historical records with orthographic variations and complex transcription details through SQL and coding agents for systematic data analysis.}
    \label{fig:challenges}
\end{figure}

This study introduces a generalizable framework that uses Large Language Models (LLMs) to assist historians in querying and analyzing historical cadastral data. Our objectives are threefold:

\begin{itemize}[noitemsep]
    \item \textbf{To overcome the structural heterogeneity of historical cadastral sources}, which include orthographic variations, transcription inconsistencies, and non-standardized formats that hinder large-scale analysis (see Figure~\ref{fig:challenges});
    \item \textbf{To enable domain experts to interact with these datasets using natural language}, by processing research questions as executable programs that retrieve, aggregate, or analyze the relevant data;
    \item \textbf{To evaluate the reliability, consistency, and scope of LLM-based agents} when applied to diachronic and spatially grounded historical records.
\end{itemize}

\noindent Our main contributions are:
\begin{itemize}[noitemsep]
    \item \textbf{A typology of historical research questions} tailored to cadastral analysis, which distinguishes between browsing (structured lookups) and prompting (contextual or multi-dataset queries), and maps them to the appropriate processing method.
    \item \textbf{A dual-agent system for historical data analysis}, combining two complementary strategies: 
    \begin{itemize}[noitemsep]
        \item A SQL agent for structured, idiographic queries;
        \item A Coding agent for more complex analytical operations, such as spatiotemporal comparisons and pattern detection.
    \end{itemize}
    \item \textbf{A scalable and interpretable workflow} that outputs verifiable code. This ensures that each answer is grounded in the source data and allows researchers to inspect or modify the generated analysis logic.
    \item \textbf{A case study on the cadastres of Venice (1740 and 1808)}, through which we evaluate our framework on real-world archival data. This includes an open benchmark of 240 expert-curated research questions (100 for structured browsing and 140 for complex prompting) that span spatial, functional, personal, and temporal analyses. The benchmark is designed to reflect authentic urban historical research and support future comparative evaluations.  Two example queries are shown in Figure~\ref{fig:challenges}.
\end{itemize}

\subsection{Related Work}

Processing historical cadastral data requires bridging challenges related to unstructured formats, orthographic inconsistency, and diachronic language variation. Existing computational strategies fall broadly into three categories: supervised machine learning models, rule-based methods, and more recently, LLM-based code generation.\\

\noindent \textit{Machine learning approaches} have leveraged advances in tabular modeling to support structured data extraction and reasoning. Models such as TabularNet~\citep{du2021tabular} and TableFormer~\citep{yang-etal-2022-tableformer} integrate neural architectures for improved table parsing, while STab~\citep{hajiramezanali2022stab} introduces self-supervised learning for diverse tabular data. Other methods, such as those presented in~\citep{zhang2024mixedtype}, aim to reconstruct incomplete historical records through data synthesis. However, these approaches generally require large labeled datasets and may struggle with the variability of historical formats.\\

\noindent \textit{Rule-based methods} are often designed with an assumed underlying relational structure~\citep{SHIGAROV2017123}, and rely on predefined rules for normalizing toponyms~\citep{garbin2005disambiguating}, recognizing descriptions, and applying historical ontologies. While interpretable and often robust in controlled settings, they tend to be brittle when faced with noisy, ambiguous, or non-standard historical data.\\

\noindent \textit{LLM-based code generation} represents a more flexible alternative. Instead of relying on fixed schema or extensive training, these systems generate executable code—typically SQL for structured lookups or Python for custom analyses—directly from user queries. This paradigm has multiple advantages: (1) it eliminates the need for extensive labeled training data, (2) it produces verifiable outputs through executable programs, and (3) it scales across a wide range of query complexities, from simple lookups to advanced spatial-temporal operations. Building on recent advances in LLM-based data agents, our framework adapts and extends developments in \textbf{(1) robustness}, \textbf{(2) flexibility}, \textbf{(3) modularity}, and \textbf{(4) adaptability} to the specific challenges of historical cadastral analysis. 

\textbf{(1) Robustness} is addressed through \texttt{InfiAgent-DABench}~\citep{hu_infiagent-dabench_2024}, which formalizes benchmarks for execution consistency—defined as the statistical stability of results from repeated executions of equivalent queries. We adopt this methodology to quantify run-to-run variance as a proxy for output reliability. 

\textbf{(2) Flexibility} is tackled by \texttt{OpenAgent}~\citep{xie_openagents_2023}, which integrates SQL and Python execution within a single agentic framework, enabling seamless transitions between structured retrieval and procedural computation. While this architecture informed our design, our experiments confirm that SQL alone is ill-suited to the irregular schemas and semantic variability characteristic of cadastral sources. Kapoor et al.~\citep{kapoor_ai_2024} further propose techniques for improving cost efficiency and reducing debugging iterations, which have guided our implementation choices. 

\textbf{(3) Modularity} is exemplified by \texttt{CodeChain}~\citep{le_codechain_2024}, which promotes reusable, composable analysis components—an essential property when applying similar transformations across heterogeneous datasets. 

\textbf{(4) Adaptability} is central to \texttt{KwaiAgents}~\citep{pan_kwaiagents_2024}, whose architecture combines planning, memory, and tool use to support dynamic task execution in varied contexts. We draw on this principle to develop mechanisms that select and combine spatial, temporal, and semantic analyses according to query context and dataset properties as in Majumder et al.~\citep{majumder_data-driven_2024}.

\section{Cadastral Data of the City of Venice}

The 1740 Catastico represents a \textbf{textual} survey system managed by the Collegio dei Dieci Savi in Rialto, designed to administer the Venetian tithe, a 10\% property tax introduced in 1463. Property owners submitted self-declared 'Condizioni di Decima' or 'Polizze' which detailed property type, location, status, and income. These submissions were organized by district and sequentially numbered based on submission order, with taxation calculated from declared rents \citep{chauvard_catastici_2007}. The general information structure of the document is displayed in Figure~\ref{fig:info_structure}a. Following a major archival fire in 1514, redecimation efforts occurred sporadically, with significant collections in 1514, 1537, 1566, 1581, 1661, 1711, and 1740. Unlike later geometric cadastres (e.g., the first one introduced in Venice in 1808), the Catastico did not integrate cartographic representation. Instead, records were generated through door-to-door surveys conducted by censors, who documented owners, tenants, toponyms, property functions, and rents (see Fig. 2-A).

\begin{figure}[!h]
    \centering
    \includegraphics[width=.85\linewidth]{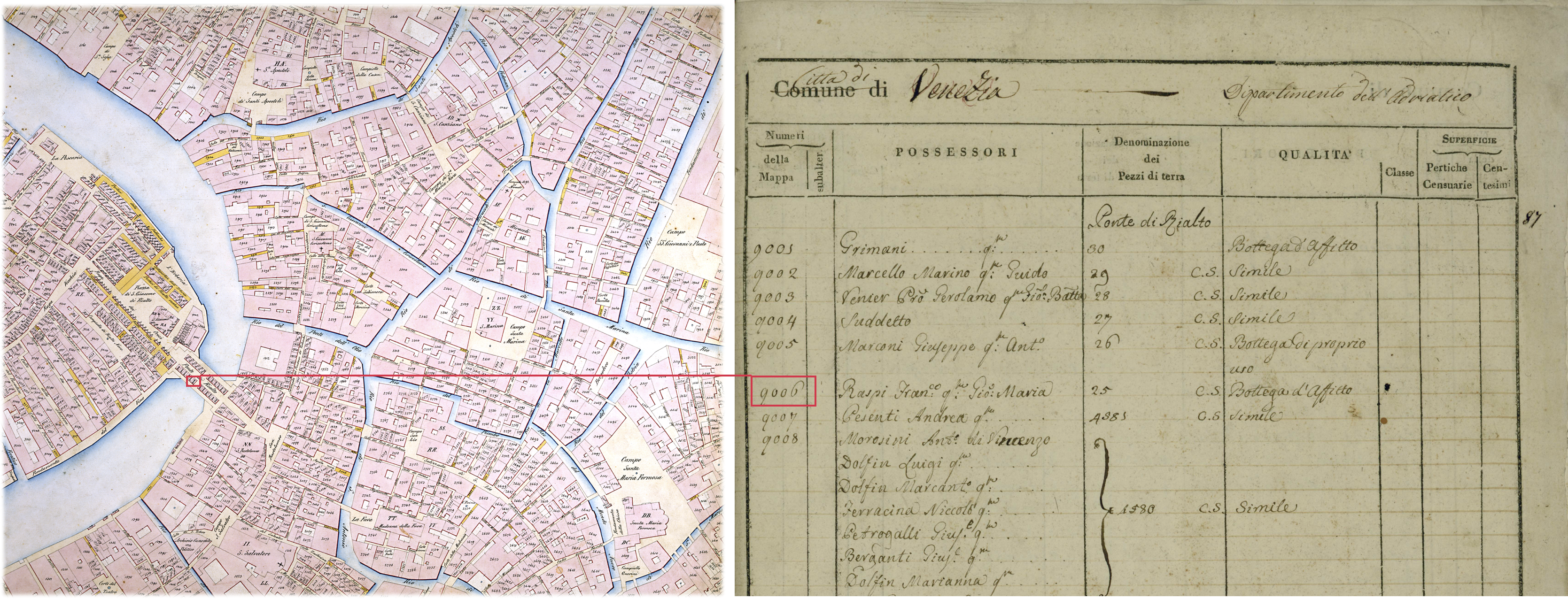}
    \caption{\textbf{The dual information system of the 1808 cadaster.} Each parcel mention in the textual document is geolocalized on the cadastral map through the same ID code}
    \label{fig:dual_system}
\end{figure}

In contrast, the first \textbf{geometric} cadastre in Venice was established in 1808~\citep{pavanello_i_1981}, adhering to French administrative standards \citep{clergeot_recueil_2008}. As displayed in Figure~\ref{fig:dual_system}, it operates as a dual system, combining cartographic maps of parcels with textual records that document ownership, location, function, and area. Each parcel is assigned a unique number, which is cross-referenced with the records that catalog owners, toponyms, uses, and dimensions (see Figure~\ref{fig:info_structure}b). Ownership records include individual names, family relationships, and institutions, while functions are classified using a codified Italian ontology from 1808 \citep{di_lenardo_approche_2021}. In particular, the terms reflect historical usage; for instance, a “shop” is referred to as 'bottega', rather than the modern Italian term 'negozio'.

These systems reflect complementary approaches to surveying. The 1808 cadastre is cartographic, parcel-based, and systematic, while the Catastico is textual, household-focused, and income-oriented. Despite these differences, both systems exhibit stable informational structures \citep{chauvard_catastici_2007}. Both have been digitized and transcribed: in the 1808 cadastre, parcel identifiers, codes, and toponyms were automatically transcribed and subsequently verified manually \citep{ares_oliveira_machine_2017,ares_oliveira_deep_2019}. For the 1740 Catastico, geolocation was achieved by correlating toponyms with contemporary maps, reconstructing the censors’ survey paths, and identifying shared features between parcels recorded in 1740 and 1808.

\begin{figure}[!h]
    \centering
    \includegraphics[width=.7\linewidth]{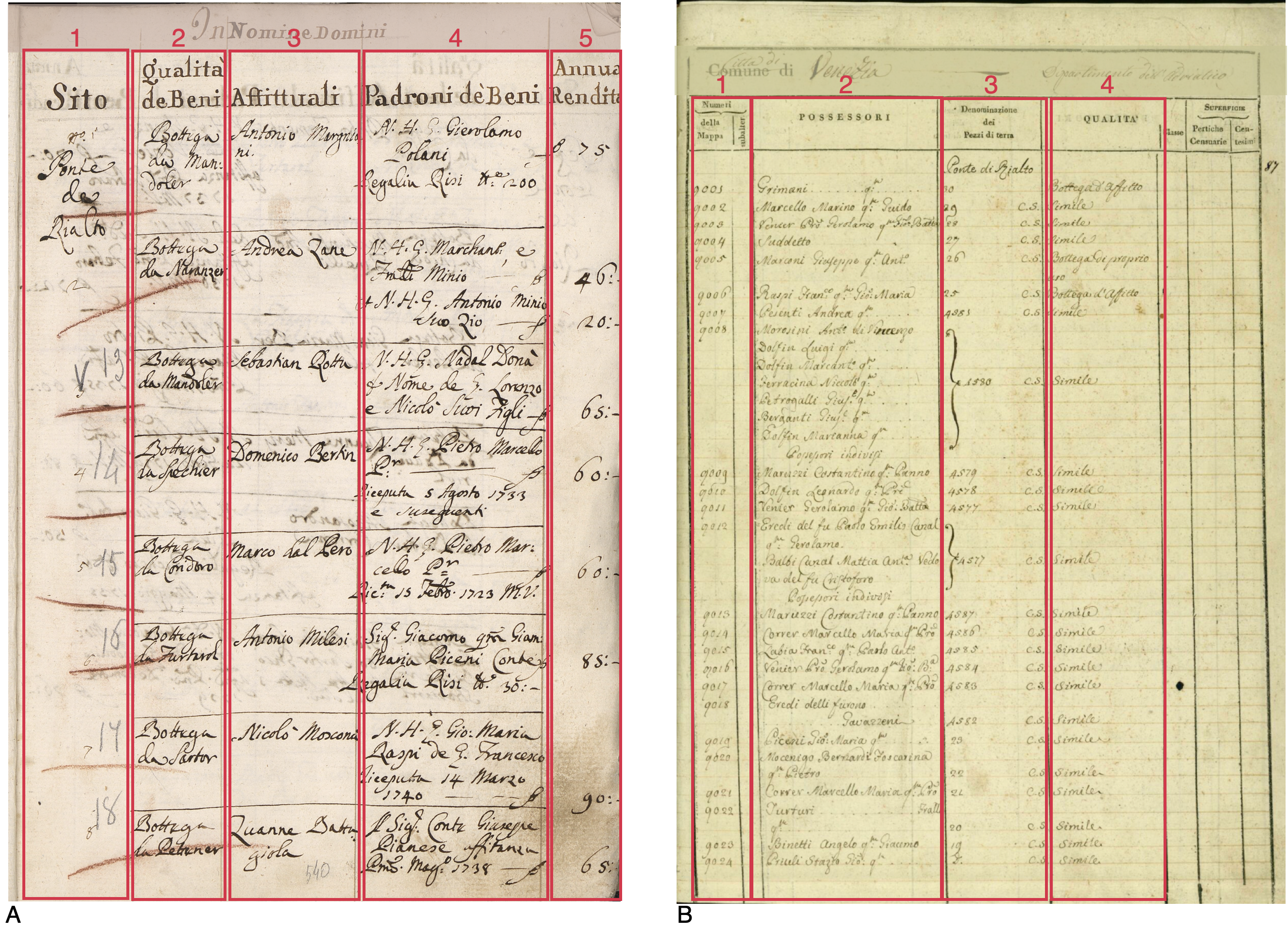}
    \caption{\textbf{The informational structure.} (A) Catastici 1740 and (B) Sommarioni 1808. The structure of the two documents is as follows for (A) : 1)place name 2)urban functions 3)tenants 4)owners 5)annual income (B) : 1)cadastral parcel identifier corresponding to a number on the map 2)owners 3)door number 4)urban functions}
    \label{fig:info_structure}
\end{figure}

Together, these datasets encompass more than 34,000 data points, providing detailed information on owner professions, tenants, property functions, rents, areas, and geolocations. In this context, a 'data point' corresponds to digitization of a single cadastral entry (row) containing all associated attributes such as parcel identifier, owner, profession, tenant and function with occasional missing values across fields.

\section{A typology of historical questions related to cadastre data}
\label{sec:typo_quest}
In urban historical research, cadastral data are used to link people and urban functions to territories~\citep{Ekamper15032010,ares_oliveira_deep_2019,Lelo_2020}. There are several objectives when consulting these records. The first is to consult them to identify the location of one or more people or urban functions in a specific place or places. The second is to investigate more general principles and test hypotheses about specific places, groups of people, types of function, or compare periods of time, particularly the past with the present. These research questions often involve complex statistical operations to aggregate and compare information that spans multiple data points from potentially different datasets. 

In our evaluation, the first type of question assesses an \llm’s ability to retrieve, combine, and compute relevant details—such as owners’ names, urban functions, and toponyms—thus using the model as a historical dataset browser. The second type probes the model’s capacity to answer more complex queries that require semantic understanding and spatio-temporal reasoning within the historical context.

\subsection{Browsing questions}
\label{subsec:browsing_questions}

Browsing questions can be categorized into 1) simple aggregation queries and 2) more complex relational queries. Simple aggregation queries focus on straightforward retrieval of information, such as calculating the total rent revenue generated by properties of a specific type. For example, a researcher might ask, ``\textit{What is the total rent revenue generated from properties of the `bottega da casarol' variety?}``. This type of query provides quick insights into financial aspects of property types, allowing immediate analysis of income generated from specific categories.

In contrast, relational queries delve deeper into the dataset to examine the relationships and patterns among various data points, such as identifying how many families own properties across multiple categories. An example of such a question could be, ``\textit{How many families own properties of more than one type category?}``. These relational queries are essential for revealing trends in property ownership and usage, providing insights into socio-economic dynamics within urban settings. Table~\ref{tab:browsing_quest_sql} displays an aggregation as well as a relational question with their related SQL queries. The set of 100 hand-crafted browsing questions about
the Catastici 1740 dataset are available in the Additional Methods.

\begin{table}[h!]
    \centering
    \begin{tabular}{p{4cm}|p{8cm}}
        
        \textbf{Browsing Question} & \textbf{SQL Query} \\ 
        \hline
        \multirow{3}{4cm}{\textit{What is the total rent revenue generated from properties of the `bottega da casarol' variety?}} & 
        \begin{lstlisting}[basicstyle=\ttfamily\scriptsize,frame=none,xleftmargin=-15pt]
        SELECT SUM("Rent_Income") 
        FROM catastici 
        WHERE "Property_Type" = "bottega da casarol";
        \end{lstlisting}  \\
        
        \multirow{4}{3cm}{\textit{How many families own properties of more than one type category?}} & 
        \begin{lstlisting}[basicstyle=\ttfamily\scriptsize,belowskip=-4pt,xleftmargin=-15pt]
        SELECT COUNT(*) 
        FROM (
            SELECT "Owner_Family_Name"
            FROM catastici
            GROUP BY "Owner_Family_Name"
            HAVING COUNT(DISTINCT "Property_Type") > 1
        ) AS families_with_multiple_types;
        \end{lstlisting} \\ 
    \end{tabular}
    \caption{\textbf{Simple (top) and relational (bottom) browsing questions and their corresponding SQL queries.}  Simple queries only require a single selection in the dataset while relational queries imply multiple selections.}
    \label{tab:browsing_quest_sql}
\end{table}

\subsection{Prompting questions}

Prompting questions are designed to go beyond mere data retrieval or the aggregation of information from individual datasets. Unlike browsing questions, which rely on exact matching of entities and are thus not robust to typos, synonyms, or variations -- and require users to know precisely which data points exist within the dataset-- prompting questions aim to leverage multiple data sources along with common-sense understanding to uncover richer, more nuanced insights. Such questions require a deep understanding of linguistic subtleties, particularly in categories such as professions, ownership, and the intricate interrelations among entities within tabular data. This involves not only extracting explicit information, but also interpreting implicit connections. Furthermore, prompting questions frequently demands a conceptual grasp of spatial and temporal dynamics to effectively organize and contextualize data. In certain instances, city-specific knowledge becomes crucial for identifying diachronic language, local customs, or accurately inferring distances.

After careful analysis of the datasets and their potential applications, we identified that meaningful questions about historical cadastral data could be organized into four distinct categories. The first category leverages the geocoordinates of the cadastre entries to examine spatial distributions, enabling queries that bridge past and present urban landscapes. By relating historical properties to relatively stable urban landmarks such as churches and squares (extracted from OpenStreetMap~\citep{OpenStreetMap}), we can investigate how individuals and properties were allocated across diverse areas. Although these landmarks may have undergone modifications or reconstructions over time, they often maintain their general location and social function, serving as semipersistent spatial anchors for historical analysis. The second category is dedicated to building functions, exploring the intended purposes or uses of various structures within the urban environment. The third category focuses on personal information, examining demographic and socioeconomic characteristics associated with individuals in the cadastral data. Finally, the fourth category targets temporal analysis, specifically comparing data over two distinct periods to reveal trends, shifts, or patterns over time.

\begin{table}[!h]
    \centering
    \resizebox{.95\columnwidth}{!}{%
    \begin{tabular}{p{9cm} c c}
         \hline
         \textbf{Questions} & \textbf{Category} & \textbf{Answer Type} \\
         \hline
         \textit{Is the building rent price correlated with the distance from the nearest square in 1740?} & Spatial & Yes/No \\
         \textit{Which neighborhood contains the buildings with the highest rent price on average in 1740?} & Spatial & Entity \\
         \textit{How many people live within 100 meters from the closest church in 1808?} & Spatial & Number \\
         \textit{On average, are buildings with multiple functions more expensive than the ones with a single function in 1740?} & Function & Yes/No \\
         \textit{Which square is surrounded with the buildings with the most diverse functions within 100 meters in 1740?} & Function & Entity \\
         \textit{What is the average distance between workshops and their nearest church in 1740?} & Function & Number \\
         \textit{Is there any correlation between the family name and the profession in 1740?} & Personal & Yes/No \\
         \textit{In which parish do lawyers own the most number of buildings in 1740?} & Personal & Entity \\
         \textit{How many medical doctors are there in Venice in 1808?} & Personal & Number \\
         \textit{Did the number of buildings with more than one function increase over time from 1740 to 1808?} & Temporal & Yes/No \\
         \textit{Which family increased the number of buildings they owned the most from 1740 to 1808?} & Temporal & Entity \\
         \textit{How many new families appeared in Venice in 1808 that were not present in 1740?} & Temporal & Number \\
         \hline
    \end{tabular}}\\
    \caption{\textbf{Examples of prompting questions.} Alongside their category and expected output format.}
    \label{tab:questions}
\end{table}

In sum, we have curated a comprehensive set of 140 questions, which we are releasing as an open-source resource along with this paper. All questions have been validated by urban specialists to ensure their relevance for urban analysis applications. Table~\ref{tab:questions} presents a selection of questions from each category. The questions encompass diverse expected output formats, ranging from binary yes/no responses to numerical values or the identification of specific entities.

Unlike browsing questions, designing SQL queries for prompting questions presents significant challenges due to their inherently complex nature. Prompting questions often require insights that extend beyond the information readily available in the dataset's structure or columns. The intricate operations necessary for these questions move beyond simple data filtering and aggregation, involving advanced processes such as semantic searches, spatial computations, and statistical tests or correlation evaluations. Additionally, certain prompting questions demand the incorporation of external knowledge or common-sense reasoning, which cannot be encapsulated purely through SQL. 

\section{Methods}
\label{sec:methods}

\subsection{Overview}

Our framework combines two complementary approaches, each tailored to a category of research questions defined in Section~\ref{sec:typo_quest}:
\begin{enumerate}[noitemsep]
    \item \textbf{SQL Agent} --- designed for \textit{browsing} questions that involve structured lookups and relational queries over a single historical dataset.
    \begin{itemize}[noitemsep, topsep=0pt]
        \item \textbf{Input:} a natural language question.
        \item \textbf{Output:} an SQL query executed on the target table to return results.
    \end{itemize}
    \item \textbf{Coding Agent} --- designed for \textit{prompting} questions requiring integration of multiple datasets and complex operations, such as spatial, temporal, or semantic reasoning.
    \begin{itemize}[noitemsep,topsep=0pt]
        \item \textbf{Input:} a natural language question.
        \item \textbf{Output:} an executable Python program that performs the required analysis, optionally including multi-step planning, entity extraction, and consistency evaluation.
    \end{itemize}
\end{enumerate}
While both agents generate executable code directly from natural language, their architectures differ: the SQL Agent uses a single text-to-SQL model over a defined schema, whereas the coding agent coordinates multiple specialized LLM components to extract entities, plan analyses, and generate Python code.

\subsection{Technical Data Representation}
\label{sec:technical_data}

Our evaluation uses two historical cadastral datasets and a modern geographic reference dataset, represented in a simplified, analysis-ready form for the purposes of this study.

\begin{itemize}[noitemsep]
    \item \textbf{Catastici (1740)} --- a tabular dataset with seven columns: 
    \texttt{Catastici\_ID} [integer], 
    \texttt{Owner\_ID} [integer], 
    \texttt{Owner\_First\_Name} [text], 
    \texttt{Owner\_Family\_Name} [text], 
    \texttt{Property\_Type} [text], 
    \texttt{Rent\_Income} [integer], 
    \texttt{Property\_Location} [text].

    \item \textbf{Sommarioni (1808)} --- a comparable cadastral dataset with owner and property attributes, structured for cross-temporal comparison with the Catastici.

    \item \textbf{Landmarks dataset} --- a set of semi-persistent urban features (e.g., churches, squares) extracted from OpenStreetMap, represented as point geometries. These landmarks serve as spatial anchors for contextualizing historical property locations within a stable geographic framework.
\end{itemize}

\subsection{SQL Agent}
\label{sec:sql_agent}

\textbf{Questions.}  
We curated 100 expert-designed browsing questions on the Catastici (1740) dataset, covering simple retrieval and relational queries, as detailed in the previous section\\

\begin{figure}[!h]
    \centering
    \includegraphics[width=0.85\linewidth]{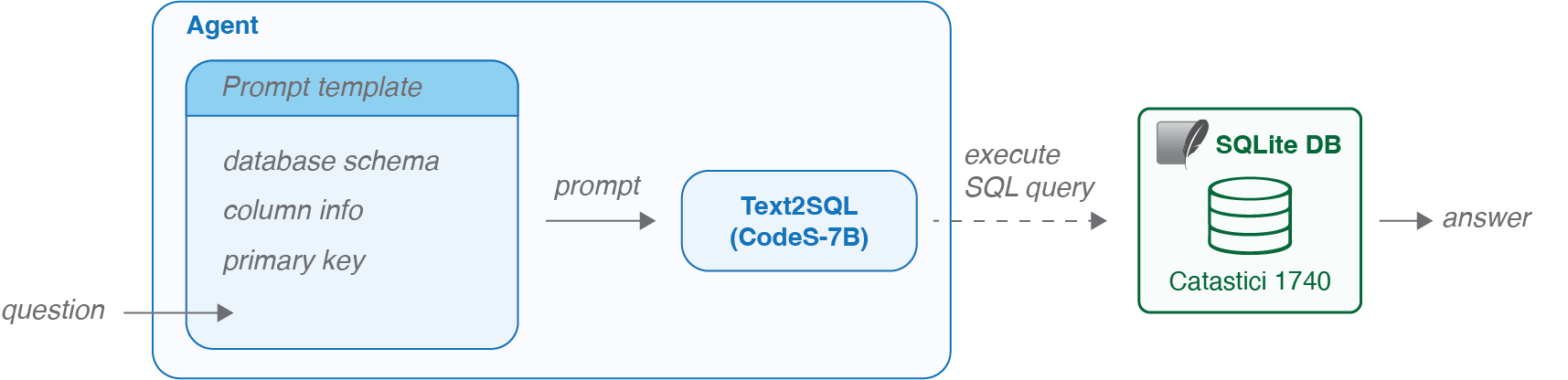}
    \caption{\textbf{The SQL-Agent.} Questions are fed to the system into a prompt engineered to match with the \texttt{CodeS} model requirements. }
    \label{fig:sql-agent}
\end{figure}

\noindent \textbf{Architecture.}  
Natural language questions are converted into SQL queries using the open-source text-to-SQL model \texttt{CodeS-7B}~\citep{li_codes_2024}. Each prompt includes a detailed description of the table schema and its columns. We evaluate two prompting settings: in the zero-shot setting, the question is provided with table metadata only; in the three-shot setting, it is provided alongside three example question–query pairs. For each question, the model is run four times with identical inputs, and the final SQL query is selected by majority voting before being executed via SQLite to obtain the corresponding answer. Figure~\ref{fig:sql-agent} provides an illustration of the SQL Agent. For further clarity, examples of the prompts designed to interact with the \texttt{CodeS-7B} model are provided in Appendix~\ref{prompt:sql}.

\subsection{Coding Agents}
\label{sec:prompting}

\begin{figure}[!h]
    \centering
    \includegraphics[width=0.85\linewidth]{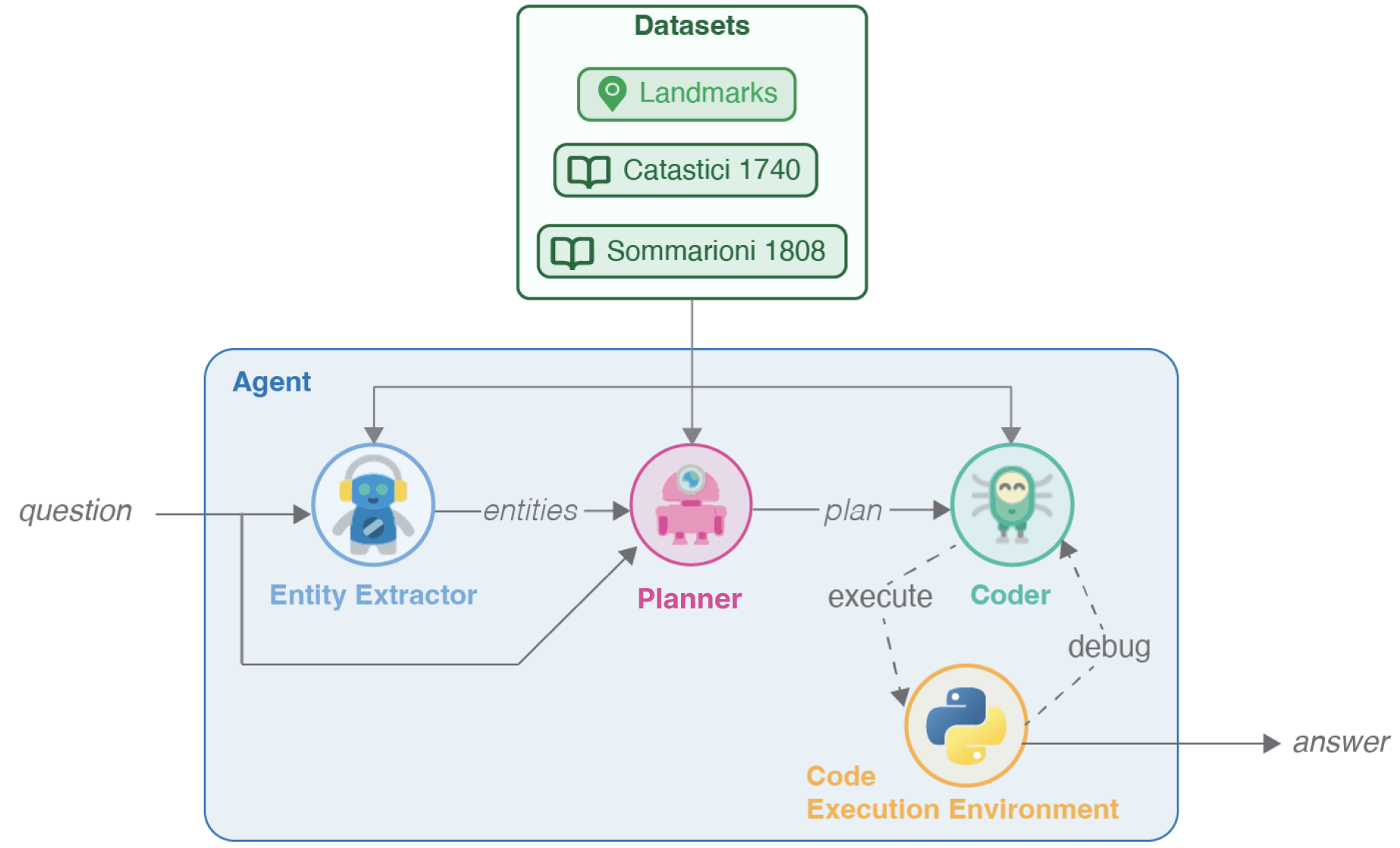}
    \caption{\textbf{The coding agent.} The agent receives a question and consults different datasets to 1) extract the entities being referred to; 2) creates a plan to answer it; and 3) produces and runs a python script to generate an answer.}
    \label{fig:citicad}
\end{figure}

\noindent \textbf{Questions.} As discussed in Section~\ref{sec:typo_quest}, prompting questions fall into four categories: spatial, functional, personal, and temporal. Spatial questions locate and organize records from the Catastici~(1740) and Sommarioni~(1808) datasets relative to landmarks from the OpenStreetMap-based Landmarks dataset (Figure~\ref{fig:citicad}), while temporal questions identify and compare patterns across the two periods. The coding agent integrates these three datasets, using semi-persistent features such as churches and squares as spatial anchors. Although these structures may have changed physically, their stable locations and enduring social functions provide a consistent framework for situating historical cadastral data within contemporary geography.

\noindent \textbf{Architecture.}  
The coding agent operates as a dialogue among three specialized components: the \textit{entity extractor}, the \textit{planner}, and the \textit{coder}. Each is an \llm prompted for a specific role (see Additional Methods), with outputs from one component guiding the next. The overall information flow is orchestrated using LangChain~\citep{langchain2022} (Figure~\ref{fig:citicad}).\\

\begin{figure}[!h]
    \centering
    \includegraphics[width=0.9\linewidth]{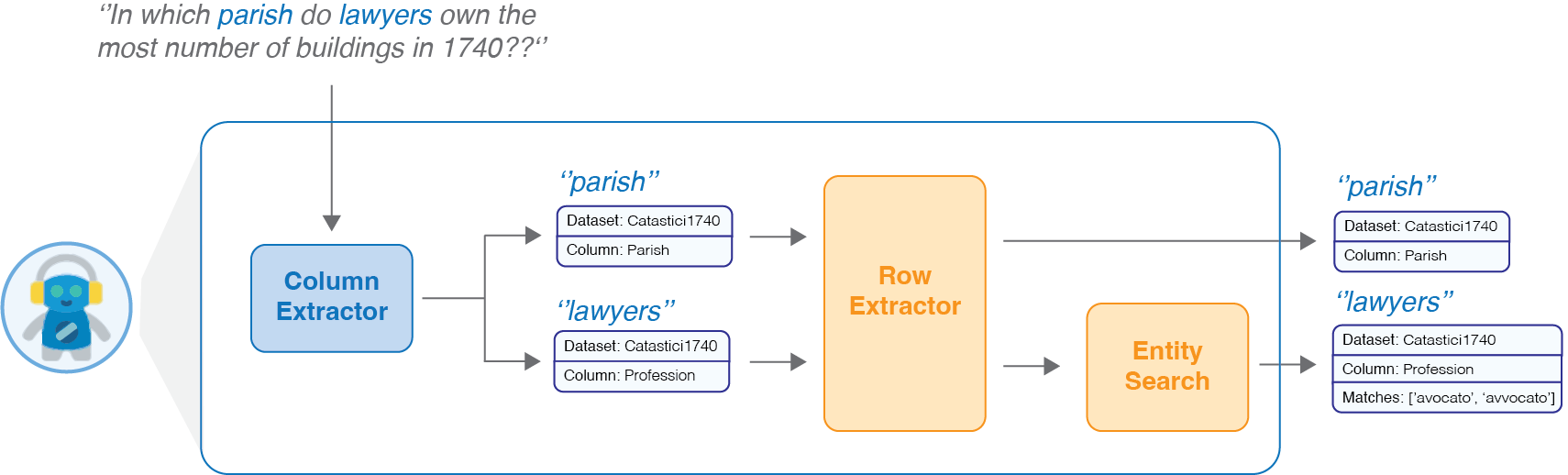}
    \caption{The Entity Extractor phase. Given a question, in this phase, we extract the most relevant rows from the datasets.}
    \label{fig:entity_extractor}
\end{figure}

\noindent \textit{Entity extraction.}
We use a Retrieval-Augmented Generation (RAG) approach~\citep{10.5555/3495724.3496517} to align broad queries with dataset content. The system selects the relevant dataset, maps question terms to columns, and retrieves entries using exact matching (e.g., \textit{avocato}), fuzzy matching for spelling variants (e.g., \textit{avvocato}), and semantic matching for related terms (e.g., \textit{procuratore}). Exact and fuzzy matching offer greater precision, while semantic search increases recall at the cost of specificity (Figure~\ref{fig:entity_extractor}).\\

\noindent \textit{Code generation.}  
Following the \texttt{Plan-and-Solve} approach~\citep{wang-etal-2023-plan}, the planner produces a detailed execution strategy based on dataset metadata, extracted entities, and query mappings. The coder then translates this plan into Python code, which is executed in a controlled environment. Errors trigger iterative debugging up to a fixed retry limit, after which unresolved queries are marked as unanswerable.\\

\noindent \textit{Consistency measures.}  
We assess \textit{Execution Consistency} (EC) as the proportion of identical outputs obtained when repeating the same question three times under different random seeds, capturing the stability of generated code and results. Following prior work~\cite{kapoor_ai_2024,hu_infiagent-dabench_2024}, we use three runs as a cost–accuracy trade-off: it is sufficient to reveal stability patterns while keeping evaluation costs manageable.

\section{Results}

\subsection{SQL Agent Results}
\label{sec:sql_results}

We evaluated the SQL Agent on 100 curated browsing questions covering retrieval and relational operations over the Catastici~(1740) dataset. Performance was assessed using two complementary metrics: \textit{exact match accuracy}, measuring the correspondence between generated and ground-truth SQL queries, and \textit{unigram overlap}, capturing lexical similarity and accounting for cases in which different query formulations yield equivalent outputs. Ground truth queries were manually authored and executed to obtain reference results. No SQL runtime errors were observed in any setting.

In the zero-shot configuration, the SQL Agent achieved 52\% exact match accuracy and 86\% unigram overlap. Providing three example question–query pairs (three-shot prompting) substantially improved these scores to 79\% and 92\%, respectively, indicating strong benefits from in-context learning (Table~\ref{tab:sql_results}).

Error analysis revealed that most failures arose from misinterpreting output specifications. For example, when asked ``How many owners receive more than 100 ducati in total rent income?'', the system returned a detailed list rather than an aggregate count. Similarly, in response to ``What is the total rent income of the top 5 earners?'', the system provided individual incomes rather than a summed value. More complex analytical tasks, such as computing the ``average rent income variance across all locations'' or the ``share of income from properties labeled as `bottega da fabro','' also proved challenging, reflecting limitations in multi-step aggregation and specialized filtering.

Overall, these results demonstrate that the SQL Agent can reliably translate natural language queries into executable SQL for idiographic browsing of historical cadastral data, with high accuracy and zero execution errors. Remaining weaknesses are primarily associated with advanced aggregation logic and nuanced output formatting. These limitations emerge most clearly when questions require multi-step reasoning, integration of heterogeneous sources, or contextual interpretation—tasks that move beyond structured browsing and into the prompting category addressed by the coding agent.

\begin{table}[!h]
    \centering
    \begin{tabular}{c|cc}
        \textbf{Model} & \textbf{Exact Match} & \textbf{Unigram Overlap} \\
        \hline
        \texttt{CodeS-7B} (0-shot) & 0.52 & 0.86 \\
        \texttt{CodeS-7B} (3-shot) & 0.79 & 0.92 \\
    \end{tabular}
    \caption{\textbf{Performance of \texttt{CodeS-7B} on browsing tasks.} Exact match and unigram overlap scores for 0-shot and 3-shot settings, with zero SQL runtime errors.}
    \label{tab:sql_results}
\end{table}

\newpage

\subsection{Coding Agent Results}
\label{sec:coding_results}

\textbf{Execution consistency.}  
We assessed the coding agent using the \textit{Execution Consistency} (EC) metric, defined as the proportion of identical answers obtained when executing the same query three times with different random seeds. EC-3 denotes perfect consistency (identical answers in all three runs) and EC-2 denotes partial consistency (identical answers in two out of three runs).

\begin{figure}[!h]
    \begin{tabular}{cc}
    \includegraphics[width=.48\textwidth]{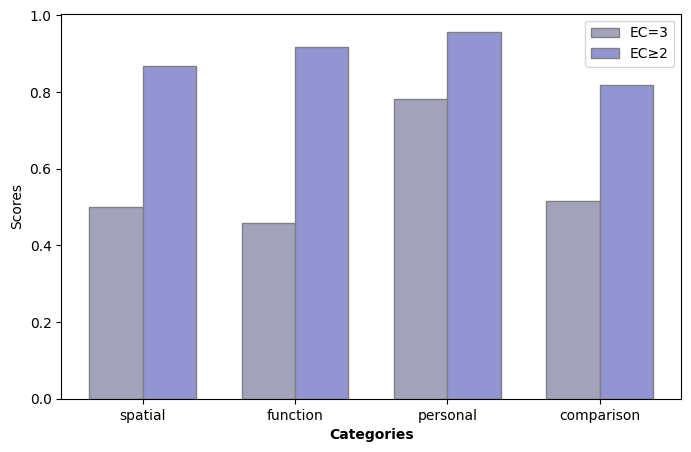} & 
    \includegraphics[width=.47\textwidth]{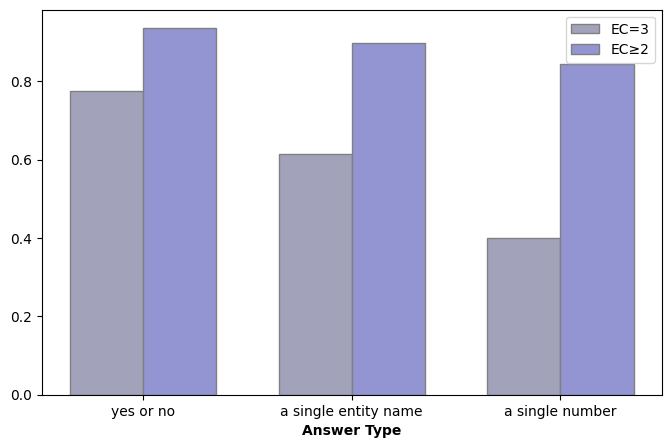} \\ 
    (a) & (b)
    \end{tabular}
    \caption{\textbf{Execution consistency (EC) of the coding agent.} (a) EC grouped by question category; (b) EC grouped by answer type.}
    \label{fig:consistency_results}
\end{figure}

Figure~\ref{fig:consistency_results} summarizes EC results by question category and by answer type. Personal and functional questions achieved the highest consistency (approximately 95\% EC-3), followed by spatial ($\approx$85\%) and comparison ($\approx$80\%) queries. Analysis by answer type revealed that yes/no and single-numerical responses were more consistent ($\approx$90\% EC-3) than entity-name extractions ($\approx$60\% EC-3), likely due to the greater determinism of binary and numerical operations compared to multi-step entity retrieval.

Manual inspection of EC-3 responses indicated that 12 of 79 consistently generated answers contained errors, yielding a 15.2\% error rate. This finding suggests that while perfect consistency is not a guarantee of correctness, it is a strong indicator of answer reliability.

\begin{figure}[!h]
    \centering
    \includegraphics[width=0.5\linewidth]{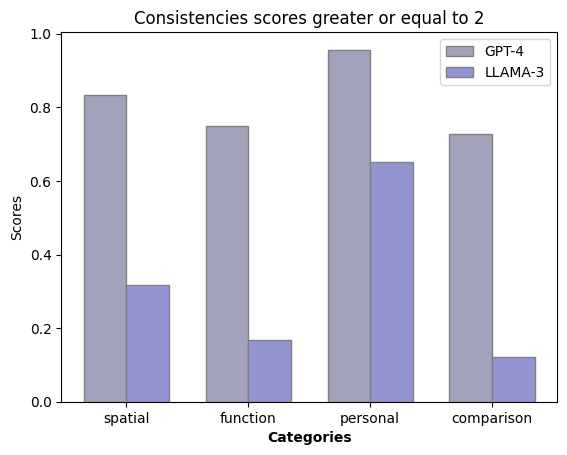}
    \caption{\textbf{Model comparison on prompting tasks.} Performance of GPT-4 and Llama-3~70B in terms of execution consistency and correctness.}
    \label{fig:gptvsllama}
\end{figure}

\textbf{Model comparison.}  
We compared GPT-4~\cite{openai2024gpt4technicalreport} and Llama-3~70B~\cite{grattafiori2024llama3herdmodels} on the same prompting tasks (Figure~\ref{fig:gptvsllama}). Both models achieved near-perfect execution accuracy (i.e., generated programs ran without error). However, GPT-4 produced markedly higher Execution Consistency and answer correctness. The primary source of this difference was not code syntax, but rather query interpretation during the planning stage. For example, when asked “Which square has the highest density of tenants?”, both models generated valid \texttt{Pandas} code, but GPT-4 correctly mapped “square” to the appropriate geographic entity type and returned campo san bartolomeo, while LLaMA misaligned the entity resolution and returned corte bollani. This illustrates that GPT-4’s advantage stems from emergent reasoning capabilities associated with larger-scale models, enabling more accurate semantic parsing of complex questions, rather than purely from code generation proficiency.

\subsection{Qualitative Analysis of Data Operations}
\label{supplsec:data_ops}

Table~\ref{tab:answer_details} illustrates how the coding agent systematically translates complex natural language questions into executable data analysis operations. Each question is decomposed into key entities and references, mapping linguistic elements (e.g., ``rent price,'' ``square'') to corresponding dataset columns and types. Once entities are identified, the agent determines the appropriate analytical procedure—such as correlation analysis, filtering, or aggregation—tailored to the query. For example, a question about the correlation between rent price and proximity to squares in 1740 is converted into a Pearson correlation calculation, while queries involving temporal comparisons or categorical relationships employ statistical methods such as chi-square tests or comparative metrics.

Two broad error classes emerged from this qualitative inspection.  
First, \textit{semantic ambiguity} in entity mapping: modern descriptors such as ``commercial building'' or ``shop'' were often mapped to historical terms like \textit{magazzeno} or \textit{locale}, which in 18th-century Venetian context usually referred to storage or ancillary spaces. This mismatch inflated counts in spatial aggregation tasks, such as identifying squares with the highest density of commercial buildings.  
Second, \textit{diachronic linguistic variation}: terms like \textit{negozio} or \textit{ufficio}, introduced in later centuries, were substituted for the historically accurate \textit{bottega} when analysing functional change over time, leading to systematic undercounting.

From the EC-3 subset of outputs, three consistent trends emerged:
\begin{itemize}
    \item \textbf{Robust mappings in well-defined domains.} Queries involving unambiguous spatial anchors (e.g., \textit{square}, \textit{church}) or numeric attributes (e.g., \textit{rent price}) were consistently mapped to correct dataset fields and processed with straightforward statistical or spatial operations, yielding high EC and correctness.
    \item \textbf{Greater variability in functional classifications.} Queries involving building \textit{function}, especially those requiring aggregation over categories (e.g., ``multi-function buildings''), were more sensitive to lexical variability and synonym coverage.
    \item \textbf{Temporal integration challenges.} Cross-dataset comparisons between 1740 and 1808 often failed to bridge lexical gaps when direct string matches were absent, highlighting the need for historical ontology alignment.
\end{itemize}

Overall, the coding agent demonstrates strong performance when entity mappings are unambiguous and align with stable schema elements. However, performance degrades when semantic drift, synonym variation, or temporal differences in terminology are present, underscoring the potential value of incorporating structured historical vocabularies into the pipeline.

\begin{table}[!h]
    \centering
    \renewcommand{\arraystretch}{1.3}
    \resizebox{\textwidth}{!}{%
    \begin{tabular}{p{6cm} p{5.5cm} p{5.5cm}}
         \hline
         \textbf{Question} & \textbf{Entity \& Reference} &  \textbf{Code Operation} \\
         \hline
         \textit{Is the building rent price correlated with the distance from the nearest square in 1740?} 
         & ``square'' $\rightarrow$ \texttt{type} in \textit{Landmarks}; ``rent price'' $\rightarrow$ \texttt{price} in \textit{Catastici\_1740} 
         & Pearson correlation between rent price and distance to nearest square \\
         \hline
         \textit{Which neighborhood contains the buildings with the highest rent price on average in 1740?} 
         & ``rent price'' $\rightarrow$ \texttt{price} in \textit{Catastici\_1740} 
         & Uses ``parish'' as ``neighborhood''; finds parish with highest average rent price \\
         \hline
         \textit{How many people live within 100 meters from the closest church in 1808?} 
         & ``Church'' $\rightarrow$ \texttt{type} in \textit{Landmarks}; ``people'' $\rightarrow$ \texttt{owner name} in \textit{Sommarioni\_1808} 
         & Filters buildings within 100 meters of a church; counts unique people \\
         \hline
         \textit{On average, are buildings with multiple functions more expensive than those with a single function in 1740?} 
         & ``single/multiple function'' $\rightarrow$ \texttt{function}; ``more expensive'' $\rightarrow$ \texttt{price} in \textit{Catastici\_1740} 
         & Compares average rent prices for single- vs. multi-function buildings \\
         \hline
         \textit{Which square is surrounded by the most diverse building functions within 100 meters in 1740?} 
         & ``square'' $\rightarrow$ \texttt{type} in \textit{Landmarks}; ``diverse function'' $\rightarrow$ \texttt{function} in \textit{Catastici\_1740} 
         & Counts building functions within 100 meters; selects square with highest diversity \\
         \hline
         \textit{What is the average distance between workshops and their nearest square in 1740?} 
         & ``workshop'' $\rightarrow$ [\textit{magazzeno}, \textit{orto}] in \texttt{function} of \textit{Catastici\_1740} 
         & Calculates average distance from workshops to nearest square \\
         \hline
         \textit{Is there a correlation between family name and profession in 1740?} 
         & ``family name'', ``profession'' $\rightarrow$ \texttt{profession} in \textit{Catastici\_1740} 
         & Creates contingency table; applies chi-square test of independence \\
         \hline
         \textit{In which parish do lawyers own the most buildings in 1740?} 
         & ``parish'' $\rightarrow$ \texttt{parish} in \textit{Catastici\_1740}; ``lawyers'' $\rightarrow$ [\textit{avocato}, \textit{avvocato}] in \texttt{profession} 
         & Counts buildings by parish; outputs parish with most lawyer-owned buildings \\
         \hline
         \textit{How many medical doctors were there in Venice in 1740?} 
         & ``medical doctor'' $\rightarrow$ \textit{medico} in \texttt{profession} of \textit{Catastici\_1740} 
         & Filters for \textit{medico}; counts number of doctors \\
         \hline
         \textit{Did the number of multi-function buildings increase from 1740 to 1808?} 
         & ``multi-function'' $\rightarrow$ \texttt{function count} in \textit{Catastici\_1740} and \textit{Sommarioni\_1808} 
         & Compares counts between 1740 and 1808 \\
         \hline
         \textit{Which family increased the number of buildings they owned the most from 1740 to 1808?} 
         & ``family'' $\rightarrow$ \texttt{name} in \textit{Catastici\_1740} and \textit{Sommarioni\_1808} 
         & Counts buildings by family in each dataset; outputs family with largest increase \\
         \hline
         \textit{How many new families appeared in Venice in 1808 that were not present in 1740?} 
         & ``new families'' $\rightarrow$ \texttt{name} in \textit{Catastici\_1740} and \textit{Sommarioni\_1808} 
         & Counts families in 1808 not present in 1740 \\
         \hline
    \end{tabular}}
    \caption{\textbf{Representative examples of EC-3 answers.} Each row links question elements to dataset columns and describes the corresponding code operation.}
    \label{tab:answer_details}
\end{table}

\section{Discussion}

Our results provide several insights into the use of LLM-based text-to-program systems for historical cadastral analysis. First, when applied to structured queries, specialized SQL and general-purpose coding agents achieve comparable accuracy: the coding agent reaches a unigram overlap of 0.85 on browsing questions, indicating that its greater versatility does not come at the expense of precision. This suggests that a single Python coding agent pipeline could address both simple and complex queries, offering a unified solution for heterogeneous analytical needs.

A central advantage of the proposed framework is its interpretability. By producing executable code rather than direct natural language answers, the system anchors every result in the source data, reducing the risk of unsupported inferences. The generated programs act as transparent reasoning traces in which analytical assumptions and methodological choices are explicitly encoded. Our comparison of GPT-4~\cite{openai2024gpt4technicalreport} and Llama-3~70B~\cite{grattafiori2024llama3herdmodels} (Figure~\ref{fig:gptvsllama}) shows a substantial performance gap, largely attributable to differences in query interpretation rather than code generation. While this underscores current limitations of open-source LLMs for complex historical analysis, rapid progress in the field suggests the gap may narrow. The framework is also city-agnostic: although demonstrated on Venetian cadastres, it can be adapted to other urban contexts with digitized records, requiring only minor schema and context adjustments.\\

Several limitations must be considered. First, the interpretability benefits of code outputs assume some programming literacy, which may limit accessibility for historians without technical backgrounds. An intermediate layer translating code into natural language explanations could broaden usability. Second, performance declines when processing diachronic linguistic variation, pointing to the need for time-aware retrieval-augmented generation methods tailored to historical dialects and evolving terminology. Third, the use of semi-permanent urban anchors (e.g., churches, squares) is effective for spatial grounding but may introduce anachronisms or oversimplifications if not critically assessed. More broadly, results must be interpreted in light of the assumptions and biases inherent in the underlying datasets.

These findings highlight promising directions for future work: improving accessibility through natural language code explanations, enhancing historical language processing, and developing more nuanced temporal–spatial integration methods. Despite current challenges, the proposed framework demonstrates that LLM-based agents can support rigorous, verifiable, and scalable historical urban research.

\paragraph{Author Contributions}

\begin{itemize}[nosep, topsep=0pt]
  \item \textbf{Conceptualization:} I.~Di~Lenardo; F.~Kaplan
  \item \textbf{Methodology:} T.~Karch; J.~Saydaliev; I.~Di~Lenardo
  \item \textbf{Supervision:} T.~Karch; I.~Di~Lenardo; F.~Kaplan
  \item \textbf{Investigation:} T.~Karch, J.~Saydaliev
  \item \textbf{Data curation:} J.~Saydaliev
  \item \textbf{Visualization:} T.~Karch
  \item \textbf{Writing -- original draft:} T.~Karch
  \item \textbf{Writing -- review \& editing:} All authors
\end{itemize}

\noindent All authors approved the final submitted draft.

\paragraph{Competing Interests}
The authors declare no competing interests.

\paragraph{Data and Code Availability Statement}
An extended version of the 1808 Venetian cadastre is available through a Zenodo repository\footnotemark[1]~\citep{di_lenardo_2025_16761169}. 
The exact dataset employed in this study, together with the full set of questions and the accompanying code, are provided in the associated GitHub repository\footnotemark[2].

\footnotetext[1]{\url{https://doi.org/10.5281/zenodo.16761169}}
\footnotetext[2]{\url{https://github.com/dhlab-epfl/venice-agents}}

\paragraph{Ethical Standards}
The research meets all ethical guidelines, including adherence to the legal requirements of the study country.

%\endnote in some journals will behave like \footnote; and \printendnotes will not output anything. 
% \printendnotes

% \defbibnote{preamble}{By default, this template uses \texttt{biblatex} and adopts the Chicago referencing style. However, the journal you're submitting to may require a different reference style; specify the journal you're using with the class' \texttt{journal} option --- see lines 1--9 of \emph{sample.tex} for a list of options and instructions for selecting the journal.}

\printbibliography

\appendix

\newpage 

\section{Additional Methods}

\subsection{Questions for Browsing the Catastici 1740}
\label{app:browsing_questions}
\footnotesize
\scalebox{.64}{
\begin{tabular}{p{0.5cm}p{18cm}}

\textbf{\#} & \textbf{Question}  \\ \hline
1 & Specify all the property types that are encapsulated in the dataset.  \\
2 & How many properties are recorded in the dataset?  \\
3 & Are you able to list every property location included in the dataset?  \\
4 & How many properties have a rental income lower than 30 ducati?  \\
5 & What is the number of estates leased at a cost exceeding 38 ducati?  \\
6 & What represents the maximum rent income detailed in the dataset?  \\
7 & How much rent income do all the properties in the dataset collectively produce?  \\
8 & What is the minimal rent income figure represented in the dataset?  \\
9 & What does the mean rent income amount to when considering all properties?  \\
10 & What is the total number of distinct property owners recorded in the dataset?  \\
11 & Could you enumerate all individuals who possess ownership of properties?  \\
12 & What is the count of distinct property locations presented in the dataset?  \\
13 & What is the total number of unique locations where the properties are situated?  \\
14 & How many unique property types are represented throughout the dataset?  \\
15 & What is the total count of "casa" properties listed?  \\
16 & What is the total number of "casa" present in the dataset?  \\
17 & What is the count of properties designated as "casa"?  \\
18 & What is the count of dataset properties that are classified as types other than "altro appartamento"?  \\
19 & How many "casa in soler" properties exist in total?  \\
20 & Which property generates the least amount of rent income?  \\
21 & What is the total rent revenue generated from properties of the "bottega da casarol" variety?  \\
22 & Who are the proprietors of estates with a rent revenue of 30?  \\
23 & What is the aggregate number of properties that the "rissardi" family members possess?  \\
24 & What is the average number of properties for each property type?  \\
25 & How much rent do properties on "calle delle carozze" generate altogether?  \\
26 & What is the total income from rent for all properties classified as "casa"?  \\
27 & What is the count of properties that generate a rental income surpassing the average within the dataset?  \\
28 & What is the count of properties that are being leased for an amount below the average rent income?  \\
29 & What are the various types of properties located on "calle della torre"?  \\
30 & What is the number of properties yielding a rental income lesser than the dataset's average?  \\
31 & How much rental income do properties typically generate in "al ponte di san provolo" on average?  \\
32 & How much rent do all the properties combined on "fondamenta san domenico" generate?  \\
33 & What variety of property can be found in "teren alli gesouiti"?  \\
34 & What are the names of the families who possess land in "rio san baseggio"?  \\
35 & Who holds ownership of properties classified as "magazen due"?  \\
36 & How much is the typical rental income for various property categories?  \\
37 & Can you identify the owner whose single property generates the highest rental revenue?  \\
38 & How much is the typical rental income for properties based on their respective locations?  \\
39 & Who are the proprietors possessing real estate with rents fluctuating between 20 to 100 ducati?  \\
40 & Can you tell me about the spectrum of rents collected on "calle de franchi"?  \\
41 & What is the average rental income generated by each family?  \\
42 & How much rental income is typically received from properties located outside of "segue rughetta verso calle del fontico" on average?  \\
43 & How much income does "carlo" "gritti" generate from rent on his property?  \\
44 & Who is the top earner from a single property in terms of rent and what is the amount?  \\
45 & Who holds ownership of the estates situated at "corte de cà celsi"?  \\
46 & What variety of holdings does "marin" "bernardo" possess?  \\
47 & Could you enumerate the various types of real estate that "marin" "bernardo" possesses?  \\
48 & Which kind of property is associated with the lowest average rental earnings?  \\
49 & Can you list the various property categories under "francesco" "patarol"'s ownership?  \\
50 & What is the count of properties located on "la calle vicina al campiel dal panizza in arzere"?  \\
51 & How many categories of properties are solely found in certain locations?  \\
52 & What is the sum of rental revenue that "francesco" "mosto" collects from his entire real estate portfolio?  \\
53 & What types of properties can be found in "calle di santa cattarina principia alle fondamenta nove"?  \\
54 & Are there any estates that generate a rental income of 120 ducati?  \\
55 & How many estates are possessed by each proprietor?  \\
56 & Which properties generate a rental income exceeding 50 ducati?  \\
57 & In each location, what is the count of "casa" properties?  \\
58 & What kind of property yields the greatest amount of total rent income?  \\
59 & Based on overall rental income, which category of property is the most profitable?  \\
60 & In what area can you find the widest variety of property types?  \\
61 & Which property location boasts the widest variety of properties?  \\
62 & Can one find properties yielding rent incomes below 60 ducati?  \\
63 & What location earns the greatest amount of total rent income?  \\
64 & Identify the property location that generates the least amount in total rent income and specify the amount.  \\
65 & Who are the individuals possessing properties in several areas?  \\
66 & Is there availability of "casa" type properties located on "calle della malvasia"?  \\
67 & Who are the proprietors of a "bottega" located at "fondamenta in fazza la beccaria"?  \\
68 & How many properties out of a hundred are found in "fondamenta de carmini"?  \\
69 & What’s the minimum amount "zuane" "panizza" earns from the property at "traghetto di san silvestro"?  \\
70 & What's the typical rental income for each type of property owned by "perina" "capello"?  \\
71 & What is the count of properties classified under "casa a pepian", and how do they contribute to the overall rental revenue percentage-wise in the data?  \\
72 & In the dataset, which entity holds ownership over the largest number of properties?  \\
73 & Does "antonio" possess any real estate on "calle dei ragusei"?  \\
74 & Regarding property ownership, what is the record for the most held by one person?  \\
75 & How is the dataset broken down by property type?  \\
76 & What is the count of proprietors possessing a minimum of one "porzione di bottega" property?  \\
77 & Who is the owner with the highest number of distinct locations for their properties?  \\
78 & Who is the principal owner of the most varied property categories?  \\
79 & Is a property in "campiello della fraterna" owned by "iseppo maria" "gallo"?  \\
80 & Who holds the title as the largest landholder in "sotto le collonelle"?  \\
81 & How many properties do owners have on average in the location of "corte carli"?  \\
82 & What's the total of property owners with investments spread across several areas?  \\
83 & In "loco vicino la calle del paradiso", are there identical property types under the ownership of different families?  \\
84 & Is the property yielding the least rental income in the dataset owned by "domenico" "michiel"?  \\
85 & How does the total earnings from rent of the "casetta, vuota, di solito affittata" stand in comparison with that from different kinds of properties?  \\
86 & Concerning the dataset, can you determine the mean amount of properties attributed to every owner?  \\
87 & How much total rent income do properties with more than 10 ducati generate?  \\
88 & How many families own property in "calle corrente dell'occa"?  \\
89 & What is the lowest income of "francesco" "giustinian"?  \\
90 & For rental properties, what share of the income is derived from those labeled "bottega da fabro"?  \\
91 & How many families own properties of more than one type category?  \\
92 & How many properties are located in the top three areas with the highest total rent income?  \\
93 & How many owners have just one property in the dataset?  \\
94 & What is the total rent income of the top 5 earners?  \\
95 & How many owners receive more than 100 ducati in total rent income?  \\
96 & How many owners have more than one type of property?  \\
97 & How is rent income distributed among properties in "rio terrà"?  \\
98 & How many properties account for the lower 50\% of total rent income?  \\
99 & How many owners do not have any properties in the most populated property location?  \\
100 & What is the average rent income variance across all locations?  \\
\end{tabular}}
\normalsize

\subsection{SQL CodeS-7B Prompts}

\begin{tabular}{cc}
\begin{promptbox}[label=prompt:sql,title=Prompt for the CodeS Model,width=.7\textwidth]
database schema: \\
table catastici , columns = [ catastici.ID ( integer ) , catastici.Owner\_ID ( integer ) , catastici.Owner\_First\_Name ( text ) , catastici.Owner\_Family\_Name ( text ) , catastici.Property\_Type ( text ) , catastici.Rent\_Income ( integer ) , catastici.Property\_Location ( text )]\\
column info: \\
ID -- Primary key ; Owner\_ID -- Unique ID of each owner of the property ; Owner\_First\_Name -- First name of the owner of the property ; Owner\_Family\_Name -- Family name of the owner of the property ; Property\_Type -- Specific type of the property given in Italian ; Rent\_Income -- Rent price of the property that the owner receives as income, given in Venice ancient gold coin ducato ; Property\_Location -- Ancient approximate toponym of the property given in Italian \\
primary key : catastici.ID \\
qyestion: \\
How many properties are there with the type of "casa"?
\end{promptbox} & \begin{promptbox}[label=prompt:sql,title=3-shot template,width=.25\textwidth]
database schema: \\
colunn info\\
primary key  \\
question 1 \\
SQL Query 1 \\
database schema: \\
colunn info\\
primary key  \\
question 2 \\
SQL Query 2 \\
database schema: \\
colunn info\\
primary key  \\
question 3 \\
SQL Query 3 \\
\end{promptbox}
\end{tabular}

\subsection{Coding Agent Implementation Details}

Open AI API \footnote{\url{https://platform.openai.com/docs/overview}} Generation hyperparameters (top p, temperature, ...) are left as default and used with a random seed. The system is implemented using the langchain framework \footnote{\url{https://www.langchain.com/}}, as it allows flexibility to implement the interaction between the agents.

\subsection{Coding Agent Prompts}

\label{app:prompts_citicad}
This section contains all the prompts used at each step of the coding agent’s process. The first two prompts serve as system prompts, provided to the agent along with a description of the datasets as part of its context. The subsequent prompts align with the agent’s workflow, as shown in Figure~\ref{fig:citicad}. The process begins with extracting references, followed by entity extraction. Next, the agent generates a plan and subsequently writes the necessary code. If errors occur during code execution, the agent debugs the code based on the Python console’s error messages. In the following prompts, all inputs used to generate prompts are highlighted in blue. In context examples are given for reference and entity extraction.

\begin{promptbox}[label=prompt:python_analysis,title=Analysis System prompt,width=.98\textwidth]
You are an expert historian. You are working with 3 datasets, one detailing buildings in Venice from 1740, another one detailing buildings in Venice from 1808 and the last one listing landmarks such as churches and squares in Venice. In the Buildings datasets (1st and 2nd datasets), each row refers to a separate building, while in the Landmarks dataset (3rd dataset), each row refers to a separate landmark. 

% The datasets are as follows: \\ 
% 1. 1740 Buildings Dataset (data/buildings\_1740.csv) columns:\\ 
% - owner\_first\_name -- first name of the building owner \\ 
% - owner\_family\_name -- family name of the building owner \\ 
% - owner\_profession -- profession of the owner \\ 
% - tenant\_name -- name of the tenant in the building \\ 
% - building\_functions -- a comma separated list of the functions the building is used as. \\ 
% - rent\_price -- numerical value that refers to Italian ducats. \\ 
% - parish -- parish that the building is located at \\ 
% - building\_functions\_count -- numerical value that is the same as the length of building\_functions \\ 
% - longitude - float, longitude \\ 
% - latitude - float, latitude \\ 

% 2. 1808 Buildings Dataset (data/buildings\_1808.csv) columns:\\ 
% - owner\_first\_name -- first name of the building owner \\ 
% - owner\_family\_name -- family name of the building owner \\ 
% - building\_functions -- a list of the functions the building serves as \\ 
% - building\_functions\_count -- numerical value that is the same as the length of building\_functions \\ 
% - building\_area -- building area, in meters square. \\ 
% - district -- district that the building is located at \\ 
% - longitude - float, longitude \\ 
% - latitude - float, latitude \\ 

% 3. Landmarks Dataset (data/landmarks.csv) columns:
% - landmark\_name -- the name of the church or the square \\ 
% - landmark\_type -- either "square" or "church" \\ 
% - longitudse - float, longitude \\ 
% - latitude - float, latitude \\ 
\end{promptbox}

\begin{promptbox}[label=prompt:python_system,title=Python System prompt,width=.98\textwidth]
"You are a highly skilled Python developer with expertise in data analysis. You are working with 3 datasets, one detailing buildings in Venice from 1740, another one detailing buildings in Venice from 1808 and the last one listing landmarks such as churches and squares in Venice. In the Buildings datasets (1st and 2nd datasets), each row refers to a separate building, while in the Landmarks dataset (3rd dataset), each row refers to a separate landmark. 

\end{promptbox}

\begin{promptbox}[label=prompt:extract_reference,title=Python reference extraction prompt \textit{(prompt inputs are in \textcolor{blue}{\{blue\}})},width=.98\textwidth]

Given a question, you need to match the phrases in the question with the columns in the dataset if applicable. Only focus on the phrases that refer to one or more columns in any of the above datasets. If none of the phrases refer to a specific dataset column, return an empty list.
If question only asks about 1740, phrases should be matched to column(s) in dataset 1. If question only asks about 1808, phrases should be matched to column(s) in dataset 2. If the question asks about both datasets, phrases can be matched to column(s) in both datasets 1 and 2.
Your output should be in the format [(detected\_phrase\_1, column\_name\_1, dataset\_number\_1), (detected\_phrase\_2, column\_name\_2, dataset\_number\_2), ...]
Note that the same phrase could correspond to a column that exist in more than 1 dataset.
Note that if a phrase refers to more than one column in a single dataset, consider each column name separately.
Note that every row is about a separate building. When the questions is about a building / buildings, it is referring to the whole dataset, and not a specific column.

For example:
If the question is "Which squares are surrounded by the most diverse set of building functions from 1740?", output [("squares", "landmark\_type", 3), ("building functions", "building\_functions", 1)], since "squares" corresponds to the "landmark\_type" column in the landmarks dataset (2nd dataset), and the information about "building functions" can be found in the column "building\_functions", and the question is asking about the time 1740, thus dataset 1.

Examples:

Question: "What is the average distance to the nearest square?"
Output: [("square", "landmark\_type", 3)]

Question: "How many houses are located near Santa Maria della Salute in 1740?"
Output: [("houses", "building\_functions", 1), ("Santa Maria della Salute", "landmark\_name", 3)]

Question: "What is the average rent price of workshops in San Polo in 1808?"
Output: [("rent price", "rent\_price", 2), ("workshops", "building\_functions", 2), ("San Polo", "district", 2)]

Question: "How many families present in Venice in 1740 still exist in the 1808?"
Output: [("families", "owner\_family\_name", 1), ("families", "owner\_family\_name", 2)]

Question: "How many people live in Venice in 1808?"
Output: [("people", "owner\_first\_name", 2), ("people", "owner\_family\_name", 2)]

Please match the relevant phrases with their corresponding column names for the following question and respond, in a natural language, in the format [(detected\_phrase, column\_name, dataset\_number)].
Question: \textcolor{blue}{\{question\}}

Let's think step by step:

\end{promptbox}

\begin{promptbox}[label=prompt:extract_entity,title=Python entity extraction prompt \textit{(prompt inputs are in \textcolor{blue}{\{blue\}})},width=.98\textwidth]

You are a given a mapping between a phrase and a column of a dataset. Your task is to hypothesise if the given phrase could correspond to a specific value in the matching column depending on the definition and data type of what should be given in the columns. \\
Respond [[True]] if you think the phrase may correspond to one or more specific values in the corresponding column. \\
Respond [[False]] if you think the phrase is just referring to the corresponding column in general, not possibly not to any specific value.
Note that Dataset is referred to with its number.

For example:
If the matching is ("squares", "landmark\_type", 3), respond [[True]] as "squares" is a specific value that should be found in the column "landmark\_type". 
If the matching is ("building functions", "building\_functions", 1), respond [[False]], as "building functions" just refers to "building\_functions" column in general, and is not a specific value we are looking for. 
Give your answer between [[]], for example [[True]] or [[False]]

Examples:

Mapping: [("square", "landmark\_type", 3)]
Output: [[True]]

Mapping: [("Santa Maria della Salute", "landmark\_name", 3)]
Output: [[True]]

Mapping: [("workshops", "building\_functions", 2)]
Output: [[True]]

Mapping: [("families", "owner\_family\_name", 1)]
Output: [[False]]

Mapping: [("near houses", "building\_functions", 2)]
Output: [[True]]

Mapping: [("people", "owner\_family\_name", 2)]
Output: [[False]]

Please hypothesise, in a natural language, if the given phrase in Mapping may refer to a specific value in the corresponding column. Respond with [[True]] or [[False]].
Mapping: \textcolor{blue}{reference}

Output: \\
\end{promptbox}

\resizebox{0.85\textwidth}{!}{%
\begin{promptbox}[label=prompt:python_plan,title=Python plan prompt \textit{(prompt inputs are in \textcolor{blue}{\{blue\}})},width=.98\textwidth]
Instruction: \\
First understand the problem, and provide a step-by-step data analysis plan only in natural language to answers the question using the provided datasets. Be as clear and explicit as possible in your instructions.  \\

You are given: \\
- Question \\
- Extracted Information of Entities: This contains the dataset and the column that the entity matches to, and the corresponding exact matches found in the dataset
- References to Corresponding Dataset and Column: This contains phrases found in the question linked to the specific dataset and column \\
- Expected Answer Format: yes/no or numerical or a single textual entity name \\

Requirements:
- The final answer should be in the format of {\textcolor{blue}{answer\_format}}. 
- Use the provided entity information and datasets
- If any of the entity information or references is meaningless, ignore it.

Question:
\textcolor{blue}{\{question\}}

Extracted Information of Entities:
\textcolor{blue}{\{entities\}}

References to Corresponding Dataset and Column:
\textcolor{blue}{\{references\}}

Step by Step Plan in Natural Language:
\end{promptbox}}

\resizebox{0.9\textwidth}{!}{%
\begin{promptbox}[label=prompt:python_code,title=Python code prompt \textit{(prompt inputs are in \textcolor{blue}{\{blue\}})},width=.98\textwidth]
Instruction:\\
Your task is to generate Python code based on the provided detailed plan to answer the given question using the provided datasets.\\

Requirements:\\
- Use the necessary libraries for data analysis in Python (e.g., pandas, numpy).  \\
- The code should be well-structured, complete, and intended to be executed as a whole. \\ 
- Write your code in the most computationally efficient way \\
- Include all code in a single code block. \\
- Give your final answer in the format of \textcolor{blue}{\{answer\_format\}}. \\
- End your code by printing only the final answer strictly following this format: "[[final\_answer]]", for example: print(f"The answer is: [[{{final\_answer}}]]")\\
- Never use `exit()` function. \\

Question:
\textcolor{blue}{\{question\}}

Step-by-Step Plan:
\textcolor{blue}{\{plan\}}

Python Code:
\end{promptbox}}

\resizebox{0.9\textwidth}{!}{%
\begin{promptbox}[label=prompt:debug_prompt,title=Python debug prompt \textit{(prompt inputs are in \textcolor{blue}{\{blue\}})},width=.98\textwidth]
Instruction:
Debug and rewrite the provided Python code. The code follows the given plan to answer the given question using the given datasets, but it contains an error. Based on the error message, could you correct the code and provide a revised version?

You are given: \\
- Question \\
- Extracted Information of Entities: This contains the dataset and the column that the entity matches to, and the corresponding exact matches found in the dataset \\
- References to Corresponding Dataset and Column: This contains phrases found in the question linked to the specific dataset and column \\
- A detailed plan to write Python code that answers the question \\
- Incorrect python code that raises an error \\
- Corresponding error message \\

Requirements:
- If any of the entity information or references is meaningless, ignore it. \\
- Use the necessary libraries for data analysis in Python (e.g., pandas, numpy). \\
- The code should be well-structured, complete and intended to be executed as a whole. \\
- Write your code in the most computationally efficient way \\
- All of your code should be included in a single code block. \\
- Give your final answer in the format of \textcolor{blue}{\{answer\_format\}}. \\
- End your code by printing only the final answer strictly following this format: "[[final\_answer]]", for example: print(f"The answer is: [[{{final\_answer}}]]")\\
- Never use `exit()` function.

Question:\\
\textcolor{blue}{\{question\}}

Extracted Information of Entities:
\textcolor{blue}{\{entities\}}

References to Corresponding Dataset and Column:
\textcolor{blue}{\{references\}}

Step by Step Plan:
\textcolor{blue}{\{plan\}}

Incorrect Python Code:
\textcolor{blue}{\{code\}}

Error Message:
\textcolor{blue}{\{error\_message\}}

Corrected Python Code:
\end{promptbox}}

\newpage

\end{document}